# Poly(ionic liquid)s: platform for $CO_2$ capture and catalysis

Xianjing Zhou,[1,2] Jens Weber,[3] Jiayin Yuan*[2]

[1] Zhejiang Sci-Tech University, Department of Chemistry, 310018 Hangzhou, China
[2] Stockholm University, Department of Materials and Environmental Chemistry (MMK), Svante Arrhenius väg 16C, 10691 Stockholm, Sweden (jiayin.yuan@mmk.su.se)
[3] Hochschule Zittau/Görlitz (University of Applied Sciences), Theodor-Körner-Allee 16, 02763 Zittau, Germany

**Abstract**

Capture and conversion of $CO_2$ are of great importance for environment-friendly and sustainable development of human society. Poly(ionic liquid)s (PILs) combine some unique properties of ILs with that of polymers and are versatile materials for $CO_2$ utilization. In this contribution, we briefly outline innovative poly(ionic liquid)s emerged over the past few years, such as polytriazoliums, deep eutectic monomer (DEM) based PILs, and polyurethane PILs. Additionally, we discuss their advantages and challenges as materials for Carbon Capture and Storage (CCS), and the fixation of $CO_2$ into useful materials.

## 1. Introduction

Excessive emission of greenhouse gases, the main component of which is carbon dioxide ($CO_2$), has been considered as the major cause to global warming, ocean acidification and expansion of deserts in the subtropics [1-4]. The International Energy Agency (IEA) reported that the global atmospheric $CO_2$ concentration passed the historically high level, 400 ppm, in 2016 [5], around 40% higher than that in the mid-1800s and it is still in an average growth rate of 2 ppm/year in the last decades [6]. In this regard, global attention has cast on the development of efficient Carbon Capture and Storage (CCS) techniques as well as fixation of $CO_2$ into useful materials [7-9].

Current well-developed CCS techniques are often classified into three categories, *i.e.* post-combustion capture, oxyfuel combustion, and pre-combustion capture. Chemical absorption of $CO_2$ by aqueous amine solutions is a conventional and well-developed post-combustion capture technology but suffers from corrosion, volatility, toxicity, degradability and high energy consumption for regeneration [10-12]. Alternatively, various highly porous adsorbents, which operate mainly *via* physical adsorption, have been studied over the past two decades for potential use in pressure/temperature swing adsorption processes (PTSA). Materials under discussion include micro/meso-porous silica or zeolites [13,14], metal-organic frameworks (MOFs) [15-19], covalent organic frameworks (COFs) [20,21], carbonaceous materials [22,23], and more [24,25]. Among them, the hybrid MOFs (up to 27 wt%) and zeolites (up to 18 wt%) exhibit exceptionally high $CO_2$ uptake around room temperature and atmospheric pressure. [26,27] A still very valuable review of different material classes for $CO_2$ capture by adsorption, also with respect to technical issues, was provided by Hedin and coworkers recently [28]. Material combinations such as zeolite/activated carbon have already been implemented into pilot-scale in real power plants.[29]

Along this line, there is considerable interest in developing alternative techniques. Since Blanchard *et al.* [30] firstly reported $CO_2$ capture by ionic liquids (ILs), ILs have attracted much attention in the field of gas capture and separation. ILs carry unique properties, such as negligible vapor pressure, low flammability, high thermal stabilities, excellent gas selectivity, and tunable properties, just to name a few, which make them multifunctional [31]. However, the high viscosity and the associated relatively low $CO_2$ sorption/desorption rates of ILs [32-34] hamper their application in gas capture.

Recent success in poly(ionic liquid)s (PILs), *i.e.* the polymeric product of ILs, promotes their usage in and beyond $CO_2$ sorption due to a variety of new features of PILs in comparison to ILs [8,35-38]. PILs are composed of covalently linked IL species [31], and carry features of macromolecules, thus elegantly combining some unique properties and functions of ILs with that of polymers (*e.g.* easy processability and shape durability). Although suffering from a relatively poor capacity of $CO_2$ (generally <10 wt%) and a high cost in comparison to commercial $CO_2$ absorbents, the affinity of PILs towards $CO_2$ can be tailor-made through judicious choice of the IL groups and the polymer backbones, as well as the polymer structures [39-42]. Thus the PIL technology in $CO_2$ utilization encompasses not only $CO_2$ capture because of its scientific interest, but also the catalytic $CO_2$ activation, sensing, and conversion to value-added chemical feedstocks and high-end polymers. This contribution presents a brief overview of newly emerging PILs, and their $CO_2$ capture and catalysis from a general perspective.

## 2. Innovative Structures of PILs

By selecting different cation and anion pairs, polymer backbone and side groups, physics and chemistry of PILs could be tailored [8] [43]. A large number of cations and anions in IL chemistry can be understood as a big library of building blocks to design polymers. Cations such as imidazolium, pyridinium, pyrrolidinium, ammonium, phosphonium, guanidinium and piperidinium, and anions categorized into carboxylates, sulphonates, sulfonamide and inorganic type have been thoroughly researched. Typical chemical structures of PILs have been summarized by previous reviews [31]. However, the rich IL chemistry allows to produce constantly new PILs with the major discovery on polytriazoliums, deep eutectic monomer (DEM) based PILs, and polyurethane PILs (Fig. 1).

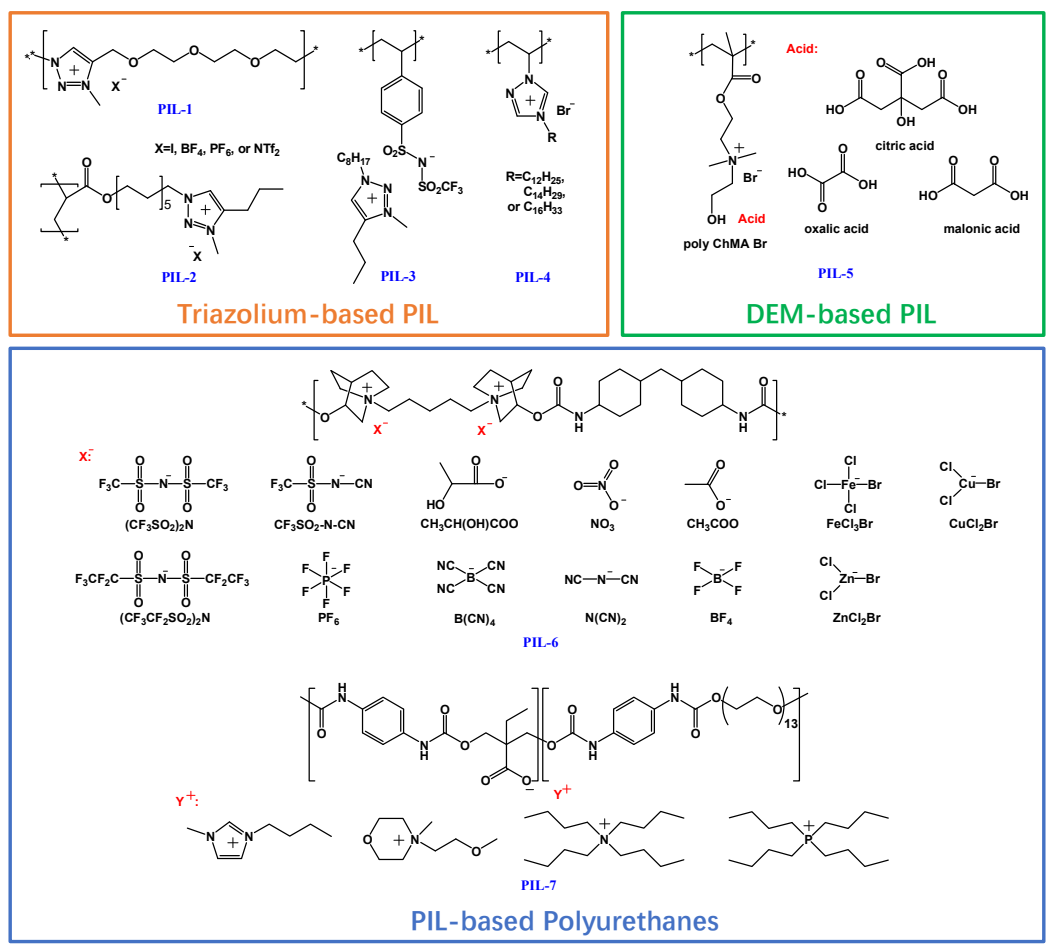

**Figure 1.** Chemical structures of PILs newly developed during the past 3 years.

Polytriazoliums have been previously seldom studied and have not been investigated in the form of PILs until very recently [44-49]. A triazolium unit has two isomers, that is, 1,2,3-, and 1,2,4-triazoliums, depending on the position of nitrogen atoms within the ring. Among them, 1,2,3-triazole derivatives that are precursors for 1,2,3-triazolium ILs can be easily obtained by copper catalyzed azide-alkyne cycloaddition [50,51]. 1,2,3-triazolium based PILs have been elegantly studied by Drockenmuller *et al*., either in the form of the main-chain (**PIL-1**) [52,53], side-chain PILs (**PIL-2**) [54-57], or as counter cations (**PIL-3**) [58]. There have been a related review published recently [46].

1,2,4-triazolium PILs are less known, though Shreeve *et al.* [59] studied briefly poly(1-vinyl-1,2,4-triazolium) as energetic materials. Miller *et al.* [60] reported a step-growth polymerization to prepare networks of a poly(1,2,4-triazolium)s by Michael addition reaction. Recently, our group constructed a series of 1-vinyl-1,2,4-triazolium-type PIL (**PIL-4**) [47-49,61].

A couple of years ago, deep eutectic solvents (DESs) emerged as non-toxic, cheap and easy-to-prepare alternatives to ILs [62]. Since then, DESs have been used as solvents, functional additives, and monomers in polymer science [63-68]. Mecerreyes *et al.* [69,70] introduced deep eutectic monomers (DEMs), which are molecular complexes formed by mixing hydrogen bond donor and acceptor. The innovative PILs based on DEMs (**PIL-5**) were obtained by photopolymerization or polycondensation.

In addition to innovative monomers, there has been active efforts to incorporate ILs into the backbone of traditional polymers, such as ionic polyurethanes. Long's group [71-73] synthesized novel cationic polyurethanes using imidazolium or phosphonium diol-based IL chain extenders. This strategy afforded polyurethane with controlled charge density normally present in the hard segments. Shaplov and Mecerreyes [74,75] synthesized a series of PIL-based polyurethanes having various diisocyanates, cations and anions (**PIL-6**). They investigated the correlation between the chemical structure and physical properties as well as the capacity of $CO_2$ capture. Colby and Einloft [76-80] have produced polyurethane anionomer with IL counter-cations (**PIL-7**). Moreover, the ILs as chain terminators [81] or cross-linkers [82] were also reported. In addition, poly(ionic liquid)s featuring thiazolium units are in general rarely reported. [83-86] In new PILs, for example, the presence of an extra more electronegative nitrogen atom in the triazolium ring in comparison with imidazolium, a variety of hydrogen bond donors in DEM-based PILs, or designable backbones of ionic polyurethanes may greatly enhance the $CO_2$ absorption capacity.

## 3. PIL for $CO_2$ capture

Some reports show that PILs exhibit a higher $CO_2$ sorption capability and faster sorption rates than the corresponding ILs [35,37,87]. The results of several groups that studied previously the structural factors in the $CO_2$ sorption capacity, including cation, anion, and polymeric backbone of PILs, are helpful to better understand this finding, [8].

Membranes are commonly used in gas separation techniques.[88-91] Gases show different permeability when they are forced to pass a membrane. The transport is usually described by a solution-diffusion mechanism and selectivity can be thought to be highly influenced by i) intermolecular interactions between the solute (gas) and the polymer matrix,[92] and ii) the (fractional) free volume of the polymer matrix. Both factors are affected by the chemical nature of the matrix and ionic liquid monomers.

In general, the type of cation plays a dorminant role in defining the PIL features in $CO_2$ sorption while in ILs anions are more important [87,93,94]. The capacities of PILs with various types of cations and anions decreased in the order of ammonium > pyridinium > phosphonium > imidazolium and $BF_4^- > PF_6^- > (CF_3SO_2)_2N^-$, according to early studies [95,96]. PILs with backbone from polystyrene (PS) to polymethylmethacrylate (PMMA) and polyethylene glycol (PEG) have a $CO_2$ sorption capacity in the order of PS > PMMA > PEG [87,93,94], an effect that partly can be ascribed to different modes of interactions and free volume distribution. The substituent of quaternized cations can also affect the $CO_2$ sorption capacity [93,97-99].

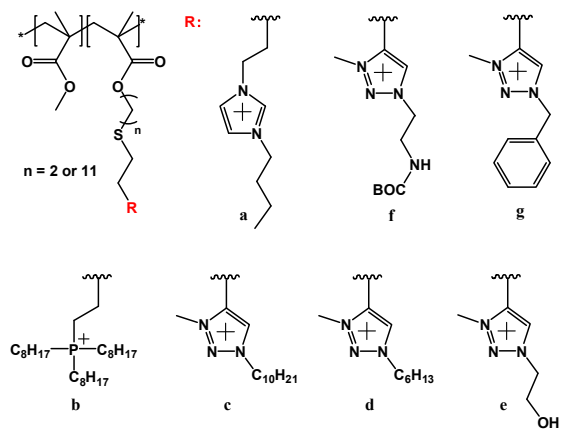

**Figure 2.** Chemical structures of PMMA-*b*-PILs block copolymers. Counter anion is Tf$_2$N. [100]

The effects of the emerging PIL structures on the sorption of CO$_2$ has been studied. Mauter *et al.* [100] synthesized a series of PMMA-*b*-PIL block copolymers (PIL-BCPs, Figure 2) with a wide range of pendant ILs including allyl functionalized imidazolium (**a**), phosphonium (**b**), and triazolium (**c-g**) cations and the Tf$_2$N anion. They pointed out that phase separation in PIL-BCPs increased CO$_2$ permeability, being the highest for PIL-based materials. The performance of CO$_2$ capture for polyDEMs (**PIL-5**, Figure 1) was investigated experimentally and computationally. The CO$_2$ sorption capacity is comparable or even higher than conventional PILs carrying fluorinated anions. The interaction of acid-containing polyDEMs with CO$_2$ follows the sequence of citric acid > oxalic acid > malonic acid [70]. For ion polyurethanes, the nature of anion, cation and the structure of the diisocyanate are all affecting the CO$_2$ uptake, although the anion factor seems to be more dominant [101]. The CO$_2$ sorption capacity of PILs based on 4,4'-methylene bis(cyclohexyl isocyanate) and diquinuclidinium as cationic backbone and 13 different anions (**PIL-6**, Figure 1) were determined at 273 K and 1 bar, which follow the order of BF$_4$ > Ac > PF$_6$ > B(CN)$_4$ > CH$_3$CH(OH)COO > NO$_3$ > (CF$_3$CF$_2$SO$_2$)$_2$N > (CF$_3$SO$_2$)$_2$N > FeCl$_3$Br> CF$_3$SO$_2$-N-CN > N(CN)$_2$ > ZnCl$_2$Br > CuCl$_2$Br [101]. Inorganic anions such as BF$_4$ and PF$_6$ were found to provide the best CO$_2$ sorption capacity, which was also observed in traditional PILs [8].

Even though some trends have been found, there is no clear prediction, whether PIL based membranes could be suitable for industrial scale processes to date. In our opinion, it will be necessary to screen the most promising membrane candidates in the near future with respect to important parameters such as permeability, mixed gas separation performance and stability of the membranes against aging and degradation. Measurement of CO$_2$ capacity alone in our opinion will not be sufficient.

Separation by selective adsorption is a second important technique next to membrane processes. Micro/mesoporous materials such as zeolites, MOFs, COFs, and porous carbons are actively investigated for their excellent CO$_2$ adsorption ability [13-15,21,22]. Porous PILs combine increased intermolecular interactions due to their ionic character together with a micro/mesoporous network. As an example, Dani *et al.* [102] reported about a porous polyimidazolium network, termed CB-PCPs obtained via a click reaction(Figure 3A, part a). It showed a better uptake of $CO_2$ at 1 bar and 273 K, but lower than that of zeolites or activated carbons. In their work, the $CO_2$ uptake depended on the nature of the anion (Figure 3A, part b) in the trend of $Tf_2N^- > PF_6^- > TfO^- > BF_4^- > Ac^-$.

Furthermore, difference between the $CO_2$ capture by means of ionic CB-PCPs and therefrom derived N-heterocyclic carbene (NHC)-bearing CB-PCPs was investigated. The NHC was obtained by deprotonation of imidazolium by a strong base (Figure 3A, part c). The adsorption mechanism of $CO_2$ with NHC is different from imidazoliums. NHC can react chemically with $CO_2$ forming imidazolium carboxylate that decomposes at ca. 80 °C to release $CO_2$ and restore carbene, a process that can be cycled, as shown in Figure 3B [103-105]. Dani *et al.* demonstrated that the $CO_2$ loading in the NHC-carrying CB-PCPs is in the same capacity range as that of the imidazolium CB-PCPs [102]. Such materials could be of high interest due to good gas selectivity, but it must be clearly said that they are to date in a premature state. Again, a validation of the materials in term of laboratory experiments that mimic real processes is needed.

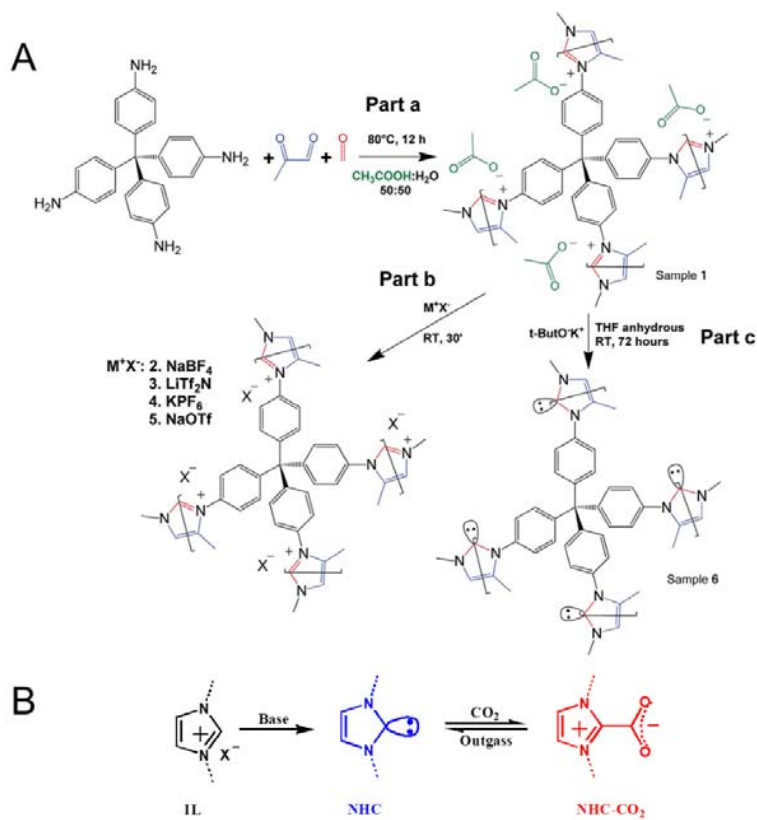

**Figure 3.** (A) The synthetic routes of ionic CB-PCPs and N-heterocyclic carbine-bearing CB-PCPs [102]. (B) Formation of NHC and NHC-$CO_2$ adducts from imidazolium-based ILs [106].

**4. PIL for $CO_2$ catalysis**

PILs can transform $CO_2$ into value-added chemical products due to their intrinsic catalytic capabilities [107]. The catalytic formation of cyclic carbonates from cycloaddition of $CO_2$ with epoxides is a classic example, reaching 100% atom economy. The catalytic cycle of ILs/PILs bearing nucleophilic counter-anions (Nu) underwent three steps, as shown in Figure 4a [108-111]: the first and the rate-determining step is the nucleophilic attack and ring-open of epoxide by an anion to form oxy-anion species (**I**); the subsequent insertion of $CO_2$ is achieved by the reaction of negatively charged oxygen atom and the electrophilic carbon atom of $CO_2$ (**II**); the formation of cyclic carbonate product after the cyclization step (**III**). Overall the catalytic study revealed that both the cation and anion of ILs/PILs affect the activity of cycloaddition. The activities of cations and anions increased in the order of imidazolium > pyridinium and $BF_4^- > Cl^- > PF_6^-$, respectively [112,113]. Moreover, Lewis acidic compounds or the hydrogen bond donors (HBDs) which resulted in the polarization of the C-O bond can facilitate the ring-open step, leading to a remarkable acceleration of the reaction rate [114-118].

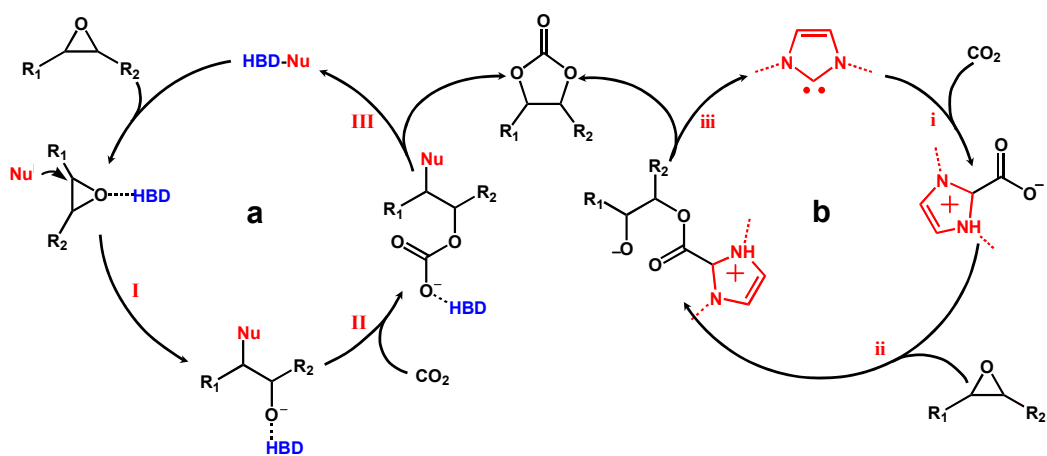

**Figure 4.** Possible mechanism for the reaction of $CO_2$ with epoxides catalyzed by PILs (a) and polyNHC adduct (b).

The first synthesis of cyclic carbonates using imidazolium-based PILs as catalyst can be dated back to 2007 [119]. A highly crosslinked PIL was prepared by copolymerization of 3-butyl-1-vinylimidazolium chloride with cross-linker divinylbenzene and showed better catalytic activity than the corresponding IL monomers and non-crosslinked PILs. Mesoporous PILs (MPILs) combine the features of mesoporous materials and ILs, representing a new direction on $CO_2$ catalysis [120-124]. Wang *et al.* [125] reported the ionothermal synthesis of a meso-/macroporous hierarchical PIL and observed its enhanced $CO_2$ conversion, which is the first metal/solvent/additive-free recyclable catalyst for heterogeneous cycloaddition of $CO_2$ at atmospheric pressure and low temperatures. The combination of porous material support and crosslinked PILs has also been studied. Ding and Jiang [126] incorporated imidazolium-based PILs into a MOF material *via in situ* polymerization of encapsulated monomers, and such material showed significantly enhanced catalytic activity under mild conditions ($CO_2$ pressure of 1 bar or lower, ≤70 °C).

The latest studies reveal that PILs could act as pre-catalysts for polyNHCs [106,127-129]. The $CO_2$ molecule could be activated by nucleophilic attack of the NHC catalyst to form zwitterionic NHC carboxylate adduct, which is more environment-tolerant than native carbenes (Figure 3B). This new feature simplifies the practical implementation of carbene-related catalytic reactions. A possible mechanism for the reaction of $CO_2$ with epoxides catalyzed by a NHC adduct runs as follows (Figure 4b) [130-133]: the zwitterionic NHC-$CO_2$ adduct firstly nucleophilic attacks epoxide to generate a new zwitterion (i); then the formed alkoxy anion nucleophilically attacks the slightly positively charged carbonyl carbon atom to produce a cyclic carbonate by intramolecular cyclic elimination (ii); finally, the released NHC reacts with $CO_2$ to regenerate the NHC-$CO_2$ adduct (iii). The development of new PILs and/or NHC complexes and the optimization of reaction conditions remains challenging. Further insight is expected from theoretical studies, which are getting more and more sophisticated. Latest results take intermolecular interactions of the surrounding into account and can help to elucidate mechanisms.[134] Finally, PILs also served as precursors for nitrogen-doped porous carbons, which showed excellent performance in $CO_2$ capture and conversion [135-138].

## 5. Conclusion

In this article, chemical structures of PILs that recently entered the field are highlighted with their utilization in $CO_2$ capture and catalysis. In addition, the $CO_2$ sorption/desorption and catalysis of deprotonated imidazolium/triazolium-based PILs, *i.e.* poly(NHC)s and poly(NHC)-$CO_2$ adducts, were also introduced. As mentioned above, the relatively poor capacity of $CO_2$ and a high cost in comparison to commercial $CO_2$ absorbents limits the practical application of PILs. Hence, future challenges are as follows according to our opinion:
   i) The development of new structures which are of general academic interest, but also of specific interest for improved efficiency in catalysis or adsorption;
   ii) Strategies for up-scaling the synthetic methods to reduce the cost of PILs;

Last but not the least, synthetic chemists are encouraged to work closely with (chemical) engineers who can provide valuable advices on the introduction of PILs-based materials into the industrial sector. The unique combination of IL properties and polymer architectures that PILs show, with all consequences and innovative potential, do render them high interest for future studies.

**Acknowledgement**
This work is supported by Strategic Research Fund at Stockholm University.